\documentclass{jfm}

\usepackage{graphicx}
\usepackage{natbib}
\usepackage[final]{changes}

\ifCUPmtlplainloaded \else
  \checkfont{eurm10}
  \iffontfound
    \IfFileExists{upmath.sty}
      {\typeout{^^JFound AMS Euler Roman fonts on the system,
                   using the 'upmath' package.^^J}%
       \usepackage{upmath}}
      {\typeout{^^JFound AMS Euler Roman fonts on the system, but you
                   don’t seem to have the}%
       \typeout{'upmath' package installed. JFM.cls can take advantage
                 of these fonts,^^Jif you use 'upmath' package.^^J}%
       \providecommand\upi{\pi}%
      }
  \else
    \providecommand\upi{\pi}%
  \fi
\fi

\ifCUPmtlplainloaded \else
  \checkfont{msam10}
  \iffontfound
    \IfFileExists{amssymb.sty}
      {\typeout{^^JFound AMS Symbol fonts on the system, using the
                'amssymb' package.^^J}%
       \usepackage{amssymb}%

      }{}
  \fi
\fi

\ifCUPmtlplainloaded \else
  \IfFileExists{amsbsy.sty}
    {\typeout{^^JFound the 'amsbsy' package on the system, using it.^^J}%
     \usepackage{amsbsy}}
    {\providecommand\boldsymbol[1]{\mbox{\boldmath $##1$}}}
\fi

\providecommand\bnabla{\boldsymbol{\nabla}}
\providecommand\bcdot{\boldsymbol{\cdot}}

\newcommand\Rey{\mbox{\textit{Re}}}  
\newcommand\Pran{\mbox{\textit{Pr}}} 
\newcommand\Ri{\mbox{\textit{Ri}}}  

\newsavebox{\astrutbox}
\sbox{\astrutbox}{\rule[-5pt]{0pt}{20pt}}

\title[Disruption of SSP/VWI states by a stable stratification]{Disruption of SSP/VWI states by a stable stratification}

\author[T. S. Eaves and C. P. Caulfield]{T. S. Eaves$^1$ \thanks{Email address for correspondence: t.s.eaves@damtp.cam.ac.uk} and C. P. Caulfield$^{2,1}$}
\affiliation{$^1$DAMTP, University of Cambridge, Cambridge, CB3 0WA, UK\\[\affilskip]
$^2$BP Institute, University of Cambridge, Cambridge, CB3 0EZ, UK}

\date{?; revised ?; accepted ?. - To be entered by editorial office}
\usepackage{epstopdf}

\begin{document}

\normalem

\maketitle

\begin{abstract}
We identify `minimal seeds' for turbulence, i.e. initial conditions of the smallest possible total perturbation energy density $E_c$ that trigger turbulence from the laminar state, in stratified plane Couette flow, the flow between two horizontal plates of separation $2H$, moving with relative velocity $2 \Delta U$, across which a constant density difference $2 \Delta \rho$ from a reference density $\rho_r$ is maintained. To find minimal seeds, we use the `direct-adjoint-looping' (DAL) method for finding nonlinear optimal perturbations that optimise the time averaged total dissipation of energy in the flow. These minimal seeds are located \replaced{adjacent}{close} to the edge manifold, the manifold in \replaced{state}{phase} space that separates trajectories which transition to turbulence from those which eventually decay to the laminar state. The edge manifold is also the stable manifold of the system's `edge state'.  Therefore,  the trajectories from the minimal seed initial conditions spend a large amount of time in the vicinity of 
some states:
the edge state; another state contained within the edge manifold; or even in dynamically slowly varying  regions of the edge manifold, allowing us to investigate the effects of a stable stratification on any coherent structures associated with such states. In unstratified plane Couette flow, these
coherent structures are manifestations of the self-sustaining process (SSP) deduced on physical grounds by \citet{Waleffe1997}, or equivalently finite Reynolds number solutions of the vortex-wave interaction (VWI) asymptotic equations initially derived mathematically by \citet{Hall1991}. The stratified coherent states we identify at moderate Reynolds number display an altered form from their unstratified counterparts for bulk Richardson numbers $\Ri_B = g  \Delta \rho H/(\rho_r  \Delta U^2) = \textit{O}(\Rey^{-1})$, and exhibit chaotic motion for larger $\Ri_B$. We demonstrate that at high Reynolds number the suppression of vertical motions by stratification strongly disrupts input from the waves to the roll velocity structures, thus preventing the waves from reinforcing the viscously decaying roll structures adequately, when $\Ri_B=\textit{O}(\Rey^{-2})$.
\end{abstract}


\section{Introduction}

\begin{figure}
	\centerline{
		\includegraphics[scale=1.1,trim=6cm 2cm 5cm 1.8cm]{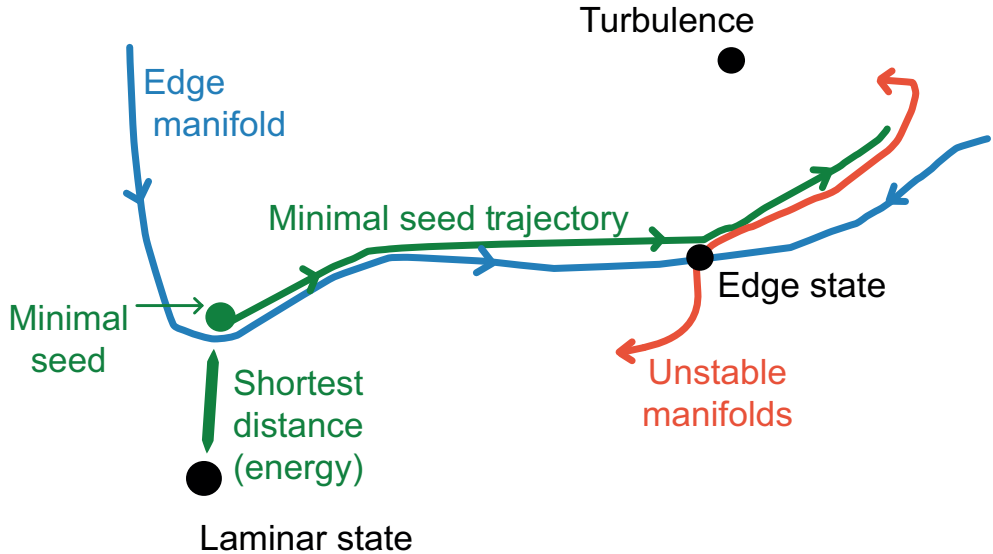}}
	\caption{Cartoon of state space showing the laminar and turbulent states, the edge manifold, edge state and its unstable manifolds, the minimal seed initial condition, and the trajectory of the minimal seed through state space.}
	\label{fig:cartoon}
\end{figure}

Stability theory has been of central importance to fluid dynamics since the pioneering work of Reynolds (1883). 
Linear stability theory is useful for determining whether or not \added{for} a given situation, \deleted{ is asymptotically stable, that is, whether }all sufficiently small perturbations\deleted{will} eventually decay to the laminar state. However, it is an inherently nonlinear question to ask what is the domain
of attraction of a known asymptotically stable solution.
The investigation of this second question is vital in order to understand why, for example, the laminar solution of plane Couette flow (PCF), i.e. the flow between two horizontal plates moving with non-zero relative velocity, is linearly stable 
at every Reynolds number, and yet sustained turbulent dynamics have been observed experimentally for Reynolds numbers as low as 325 \citep[see][]{Bottin1998}. \deleted{This indicates that }PCF \replaced{is therefore}{may be thought of as} a two state system in which both the laminar state and a chaotic turbulent state are asymptotically
attracting, and  their respective basins of attraction divide the entire state space into two distinct regions\added{, as shown schematically in figure \ref{fig:cartoon}}. These two regions are then separated by an 
edge manifold, or `edge'.\deleted{We cannot rigorously prove that PCF is a two state system, and that the concept of an edge 
manifold is well-defined, but} \replaced{T}{t}here is substantial evidence to demonstrate that for Reynolds numbers far above the critical Reynolds number for transition, below which turbulent dynamics cannot be maintained, many shear flows that have an asymptotically attracting laminar state are a two state system \citep[see][]{Duguet2008}. Near the critical Reynolds number for transition, there is growing evidence \citep[see e.g.][]{Chantry2014} that the edge manifold becomes `wrapped up' into the turbulent state, where the precise role of the edge manifold is less clear.

\deleted{Under the assumption that the system of interest has a well-defined edge manifold that separates trajectories leading to either the laminar or turbulent attractor, we must also identify what happens to trajectories that begin on the edge manifold\replaced{ must not}{Since we do not expect trajectories to} become unbounded, \added{and so}} Initial conditions on the edge manifold \replaced{are}{must be} attracted to \replaced{one of a collection of}{some} fixed point\added{s}, periodic orbit\added{s}\added{, relative periodic orbits} or chaotic attracting set\added{s} residing entirely within the edge manifold\deleted{, i.e. to an ``edge state"}. \added{The `edge state' is the state in the edge manifold which is an attractor for trajectories on the edge.} The simplest edge \replaced{state structure}{state} is a \added{single} saddle in solution space of codimension one whose stable manifold is precisely the edge manifold, and whose unstable manifold is directed into the basins of attraction of the laminar and turbulent attractors. \deleted{Whilst many edge states have been identified in various fluid dynamical systems 
, their role in the stability problem is only to provide a stable manifold that we can identify as an edge manifold and so distinguish regions of state space.}

To understand critical conditions for transition to turbulence, it is necessary to identify the smallest amplitude perturbation to the laminar state  will eventually transition to turbulence, which we call the `minimal seed', following \citet{Pringle2010}. Put in the language of edge manifolds, this is the question of what is the `closest' approach of the edge manifold to the laminar state,
as shown schematically in figure \ref{fig:cartoon}.\deleted{Since most fluid dynamical systems can experience substantial transient energy growth 
, and since an edge state in some sense represents a dynamical state with a balance between diffusion (returning to the laminar state) and instability (transitioning to turbulence), the minimal seed for turbulence is likely to be of significantly smaller energy than the edge state.} Once such a minimal seed is identified, since it lies infinitesimally above the edge manifold, and the edge manifold is the stable manifold of the edge state, its trajectory to the turbulent attractor will consist of following the edge manifold towards \replaced{a state in the edge manifold}{\replaced{an}{the} edge state} before eventually leaving the vicinity of the\deleted{edge} state along \replaced{an}{its} unstable manifold, towards the turbulent attractor. \deleted{Therefore, the edge state determines the minimal seed for turbulence in so far as it sets the shape of its own stable manifold, but this shape is not an intrinsic feature of the edge state, but rather of the equations of motion.}The edge state \added{and other states in the edge manifold} \replaced{therefore}{does however} determine a key component of the transition mechanism, since the closer an initial condition is to the edge manifold, the longer it will spend in the vicinity of \replaced{such}{edge} a state before being directed towards the turbulent attractor.

\deleted{Therefore,} \replaced{A}{a} numerical identification of an approximation to a minimal seed yields valuable stability information about the laminar state in terms of the minimal energy required for transition to turbulence, and the lack of a minimal seed for turbulence corresponds to the identification of a globally attracting laminar state. Also, the subsequent evolution of a sufficiently accurate approximation to a minimal seed should spend a large amount of time near \replaced{a state in the edge manifold}{\replaced{an}{the} edge state}\deleted{Indeed, if the approximation does not approach the edge state closely, it is still possible that the minimal seed trajectory passes through a \replaced{dynamically slowly varying}{flat} region of the edge manifold,} and have a long residency time there, before moving away from the edge manifold. \deleted{Such a \replaced{dynamically slowly varying region}{flattening} of the edge manifold and \added{its associated} long residency time of trajectories requires another coherent structure\deleted{either within or} nearby to the edge manifold in state space.}

The minimal seed 
 in unstratified PCF found by
 \citet{Rabin2012} exhibits just such a long residency time near a coherent state consisting of nearly streamwise independent rolls and streaks, before eventually transitioning to turbulence. This coherent state can clearly be interpreted as a manifestation of the `self-sustaining process' (SSP) physically described by \citet{Waleffe1997}, or equivalently as a finite Reynolds number realisation of the asymptotic vortex-wave interaction (VWI) of \citet{Hall1991} as demonstrated numerically by \citet{Hall2010}. These solutions to the Navier--Stokes equations consist of small amplitude streamwise independent roll structures creating much larger amplitude velocity streaks in the flow,\deleted{due to `lifting up' low-speed fluid from near the wall and pushing down high-speed interior fluid,} which in turn suffer instabilities to produce waves, which then reinforce the rolls, hence leading to a self-sustained process.\deleted{This process is in a delicate balance, since} \replaced{T}{t}he small amplitude roll structures are readily susceptible to viscous decay, and so the role of the waves is to inject energy into the rolls at a rate that exactly offsets this viscous decay.\deleted{In terms of the coherent state (a simple lower branch SSP/VWI state) identified by 
to which their `minimal seed' is attracted for a substantial period of its evolution, trajectories in \replaced{state}{phase} space that eventually decay or grow correspond to the waves being either too energetic, and so reinforcing the rolls too much, thus transitioning to turbulence, or being not energetic enough, in which case the perturbation slowly decays back to the laminar solution.} The SSP/VWI states are dynamically most sensitive to the feedback from the waves into the rolls since this is an inherently nonlinear process, whereas the rest of the cycle depends on linear transient growth and linear instability.

Here, we identify minimal seeds for turbulence in stratified PCF. The addition of a stable density stratification through fixing the density of the fluid at each of the horizontal plates at different (statically stable) values tends to inhibit vertical motions due to them being less energetically favourable. We\deleted{wish to} investigate how the minimal seed and its subsequent trajectory in \replaced{state}{phase} space, and any identifiable coherent structures on this trajectory, are affected by stratification and its inhibition of vertical motions. In particular, we focus on how the energy of the minimal seeds vary with increasing stratification, and how the coherent states, which in the unstratified case are SSP/VWI states, vary with increasing stratification and are disrupted due to the inhibition of vertical motions by static stability.

In section 2 we outline the algorithm used to find minimal seeds. We use the same nonlinear direct-adjoint-looping (DAL) method used by \citet{Rabin2012}, which is the nonlinear culmination of optimal perturbation analysis developed first for steady linear problems, later for time varying linear problems \citep[see][]{Schmid2007}, and most recently for fully nonlinear dynamics (see e.g. \cite{Pringle2010,Cherubini2010}, and for reviews \cite{Luchini2014,Kerswell2014}). In section 3 we show the results of the DAL method applied to stratified plane Couette flow. In section 4 we interpret the identified trajectories of minimal seeds for turbulence in stratified plane Couette flow in the language of SSP/VWI states, and demonstrate that the presence of a surprisingly weak stable stratification can still significantly modify the whole interaction by disrupting the nonlinear feedback from the waves into the roll structures, principally through an inhibition of vertical motions. We draw our conclusions in section 5.

\section{Direct-Adjoint-Looping (DAL) method}

We consider stably stratified Boussinesq PCF with a linear equation of state 
in which fluid flows between two parallel horizontal plates moving in opposite directions with relative speed $2 \Delta U$, separated by a distance $2H$, across which a constant density difference $2 \Delta \rho$ is maintained from a reference density $\rho_r \gg \Delta \rho$. Nondimensionalising with respect to $\Delta U$, $H$ and $\Delta \rho$, and decomposing the total velocity and density fields as $(\boldsymbol{u}_{tot},\rho_{tot})=(\boldsymbol{U},\bar{\rho})+(\boldsymbol{u},\rho)$, where $(\boldsymbol{U},\bar{\rho})$ is the laminar state, we obtain
\begin{eqnarray}
 \partial_t \boldsymbol{u} + (\boldsymbol{u} + \boldsymbol{U}) \bcdot \bnabla 
(\boldsymbol{u} + \boldsymbol{U}) = - \bnabla p - \Ri_B \rho \hat{\boldsymbol{y}} &+ \Rey^{-1} \nabla^2 \boldsymbol{u},\,\,\,\, (\boldsymbol{u},\rho)(y=\pm 1) = (\boldsymbol{0},0) \nonumber \\
 \partial_t \rho + (\boldsymbol{u} + \boldsymbol{U}) \bcdot \bnabla (\rho + \bar{\rho}) = 
  (\Rey \Pran)^{-1}  \nabla^2  \rho,& \,\,\,\, \bnabla \bcdot \boldsymbol{u} = 0, \,\,\,\,
  \boldsymbol{U} = y \hat{\boldsymbol{x}},\,\,\,\,
  \bar{\rho} = -y. \label{ns}
\end{eqnarray}
with the three nondimensional parameters: Reynolds number $\Rey$;
Prandtl number $\Pran$; and 
bulk Richardon number $\Ri_B$ defined
as
\begin{equation}
  \Rey = \frac{\Delta U H}{\nu} , \  \Pran = \frac{\nu}{\kappa}, 
\  \Ri_B = \frac{g \Delta \rho H}{\rho_r \Delta U^2}, \label{eq:nondim}
\end{equation}
 where $\nu$ is the kinematic viscosity and $\kappa$ is the diffusivity of density. For the calculations presented here, following \citet{Rabin2012} we choose $\Rey = 1000$, set $\Pran = 1$, and vary $\Ri_B$.

We define a \emph{nonlinear optimal perturbation} as an initial condition $(\boldsymbol{u},\rho)(t=0)=(\boldsymbol{u}_0,\rho_0)$ of given
initial total energy (kinetic energy plus potential energy) 
$E_0$, defined
as 
\begin{equation}
E_0= \left< \boldsymbol{u}_0 \bcdot \boldsymbol{u}_0 + \Ri_B \rho_0^2 \right>/2, \hbox { where } \left< \boldsymbol{a}, \boldsymbol{b} \right> \equiv \frac{1}{V} \int_V \boldsymbol{a} \bcdot \boldsymbol{b} \, \mathrm{d} V, \label{eq:e0def}
\end{equation}
 that maximises, over a given time horizon $T$, a given quantity of interest 
$\mathcal{J}(\boldsymbol{u},\rho,T)$. Here, $V$ is the 
volume of the computational domain. In order to find minimal seeds for turbulence, which are initial conditions whose trajectories eventually transition to turbulence, we typically choose an appropriately large large value\deleted{for $T$, typically} $T=300$, and we consider a functional $\mathcal{J}$ that takes heightened values in the turbulent state. With the presence of a density field, it is possible to find large amplitude waves that have an instantaneously large total energy, but are not a turbulent state. Therefore, rather than choosing the total energy density at $T$ as $\mathcal{J}$, as was done by \citet{Rabin2012} for unstratified minimal seeds, we choose the total time averaged dissipation of total energy density, the stratified generalisation of the objective functional chosen by \citet{Monokrousos2011}, for an equivalent minimal seed calculation.

We thus write\deleted{$\mathcal{J}$ as} 
\begin{equation} \mathcal{J} =
 (T \Rey)^{-1} \left[ \bnabla \boldsymbol{u} \boldsymbol{:} \bnabla \boldsymbol{u} + \Ri_B \Pran^{-1} \bnabla \rho \bcdot \bnabla \rho \right],
 \end{equation}
where $ \left[ \boldsymbol{a} , \boldsymbol{b} \right]  \equiv \int_0^T \left< \boldsymbol{a}, \boldsymbol{b} \right> \, \mathrm{d} t$,
and the maximisation of $\mathcal{J}$ can be conducted by taking variations of the augmented and constrained functional $\mathcal{L}$:
\begin{eqnarray}
	\mathcal{L} &= 
	&\mathcal{J}(\boldsymbol{u},\rho,T) 
	-\left[ \partial_t \boldsymbol{u} 
	+ N(\boldsymbol{u})+ \bnabla p + \Ri_B \rho \hat{\boldsymbol{y}}
	- \Rey^{-1} \nabla^2 \boldsymbol{u}, \boldsymbol{v} \right]  - \left[ \bnabla \bcdot \boldsymbol{u}, q \right]  \nonumber \\
	&&
	-\left[ \partial_t \rho +(\boldsymbol{u} + \boldsymbol{U})\bcdot \bnabla (\bar{\rho}+\rho) 
	- (\Rey \Pran)^{-1}\nabla^2 \rho, \eta \right] 
	+ \left< \boldsymbol{u}_0 - \boldsymbol{u}(0), \boldsymbol{v}_0 \right>\nonumber \\
	&&+ \left< \rho_0 - \rho(0), \eta_0 \right> 
	- \left(\left< \boldsymbol{u}_0, \boldsymbol{u}_0\right>/2 
	+ \left< \Ri_B \rho_0^2 \right>/2 - E_0\right)c,
\end{eqnarray}
where 
$
 N(\boldsymbol{u}) = (\boldsymbol{u} + \boldsymbol{U}) \bcdot \bnabla (\boldsymbol{u} + \boldsymbol{U}).
$
The Lagrange multipliers
$\boldsymbol{v}$, $q$ and $\eta$ are termed the adjoint velocity, pressure and density and together enforce the 
Boussinesq Navier--Stokes equations (\ref{ns}) on $\boldsymbol{u}$ and $\rho$. The initial conditions and initial energy are enforced by $\boldsymbol{v}_0$, 
$\eta_0$ and $c$. Taking variations with respect to $\boldsymbol{u}$, $\rho$, $\boldsymbol{u}_0$ and $\rho_0$ yields the following system of `adjoint' \added{or `dual'} equations that must also be
satisfied at all times and points in space by a nonlinear optimal perturbation:
\begin{eqnarray}	
	\partial_t \boldsymbol{v}
	+ N^\dagger(\boldsymbol{v},\boldsymbol{u}) + \Rey^{-1} \nabla^2 \boldsymbol{v} 
	+ \bnabla q - \eta \bnabla (\bar{\rho} + \rho) 
	-(\Rey T)^{-1} \nabla^2 \boldsymbol{u} = \boldsymbol{0}, \,\,\,\, \bnabla \bcdot \boldsymbol{v} &=& 0, \nonumber \\
	\partial_t \eta + (\boldsymbol{U}+\boldsymbol{u}) \bcdot \bnabla \eta 
	+  (\Rey\Pran)^{-1} \nabla^2 \eta 
	- \Ri_B\, \hat{\boldsymbol{y}} \bcdot \boldsymbol{v}
	-\Ri_B(\Rey \Pran T)^{-1}\nabla^2 \rho &=& 0, \,\,\,\, \label{ad} \\	
	\boldsymbol{v}(T)=\boldsymbol{0}, \,\,\,\, \eta (T) &=&0, \label{null} \\ 
         \boldsymbol{v}(0) - c \boldsymbol{u}_0 =\boldsymbol{0}, \,\,\,\,
	\,\,\,\, \eta(0) - c \rho_0 \Ri_B =0, \,\,\,\,
	\left< \boldsymbol{u}_0 , \boldsymbol{u}_0 \right> + \left< \Ri_B\rho_0^2 \right> - 2 E_0 &=& 0, \label{compat}
\end{eqnarray}
where
$
 N^\dagger(v_i,\boldsymbol{u}) = \partial_j \left((U_j+u_j) v_i\right) - v_j \partial_i (U_j+u_j).
$

The first step of the DAL method to find a nonlinear optimal perturbation is to choose an  initial condition guess $(\boldsymbol{u}_0,\rho_0)$ of energy density $E_0$. This initial condition is then integrated forwards in time to $t=T$ using the direct Boussinesq Navier--Stokes equations 
(\ref{ns}). These flow fields are stored, and used to integrate backwards in time the adjoint fields $\boldsymbol{v}$ and $\eta$ from the null `end' conditions (\ref{null}) using the adjoint equations (\ref{ad}), which actually depend on the direct fields $(\boldsymbol{u},\rho)$. Once values for the adjoint variables are obtained at $t=0$, we have compatibility conditions (\ref{compat}) relating $\boldsymbol{v}(0)$ to $\boldsymbol{u}\replaced{_0}{(0)}$ and $\eta(0)$ to $\rho_0$ that must be satisfied by a nonlinear optimal perturbation. If not satisfied, these compatibility conditions yield gradient information for the objective functional $\mathcal{J}$ with respect to changes in the initial condition $(\boldsymbol{u}_0,\rho_0)$, allowing a refined guess for the nonlinear optimal perturbation to be made. This sequence of steps is continued until convergence.

For large $T$, any initial condition that eventually becomes turbulent will be a turbulent seed  using this method. The minimal seed, however, has the special property of having the lowest critical initial energy \added{density} $E_0=E_c$ of all such seeds. We first identify
a perturbation that causes turbulence, with initial energy  \added{density} $E_0 = E_\alpha \gg E_c$. This initial condition is then used as an 
initial guess
for the DAL method at a smaller initial energy \added{density} $E_0=E_\beta < E_\alpha$, with uniformly rescaled energy. The DAL method then 
finds a more
efficient route to turbulence at $E_0=E_\beta$. This more efficient initial condition is then uniformly rescaled to
have energy \added{density} $E_0=E_\gamma<E_\beta$, and the process is repeated. This `laddering down' is continued until $E_0=E_c$, at which point any further reduction in $E_0$ cannot find an initial 
condition leading to turbulence. 



\section{Stratified minimal seeds for turbulence}

\begin{table}
  \begin{center}
\def~{\hphantom{0}}
  \begin{tabular}{ccc}
      $\Ri_B$  & $E_c$ (N) & $E_c$ (W) \\[3pt]
       0   & $2.225 \times 10^{-6} < E_c < 2.250 \times 10^{-6}$ & $8.925 \times 10^{-7} < E_c < 8.950 \times 10^{-7}$ \\
       $10^{-4}$ & $2.250 \times 10^{-6} < E_c < 2.275 \times 10^{-6}$ & N/A \\
       $10^{-3}$   & $2.600 \times 10^{-6} < E_c < 2.625 \times 10^{-6}$ & N/A \\
       $3 \times 10^{-3}$  & $3.450 \times 10^{-6} < E_c < 3.475 \times 10^{-6}$ & $1.450 \times 10^{-6} < E_c < 1.475 \times 10^{-6}$ \\
       $10^{-2}$  & $6.300 \times 10^{-6} < E_c < 6.400 \times 10^{-6}$ & $2.450 \times 10^{-6} < E_c < 2.575 \times 10^{-6}$ \\
  \end{tabular}
  \caption{Values of the critical energy \added{density} $E_c$, the energy of the minimal seed, for various bulk Richardson numbers $\Ri_B$ \added{in the two geometries N and W}. The upper bound corresponds to the flow evolutions shown in subsequent figures, and in the supplementary materials. The lower bound corresponds to an $E_0$ at which a turbulent state cannot be attained.}
  \label{tab:ec}
  \end{center}
\end{table}

\begin{figure}
	\centerline{
		\includegraphics[scale=0.4,trim=2cm 0.5cm 1cm 0cm]{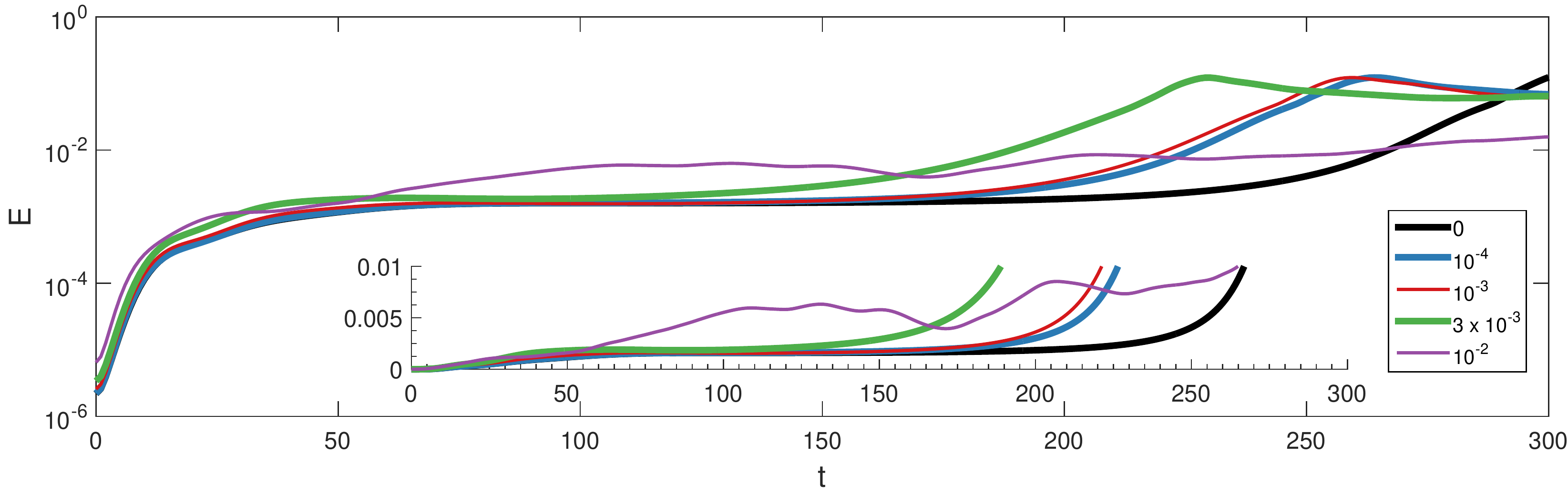}}
	\caption{Time variation of energy density $E(t)$ as defined in (\ref{eq:edef})  for the minimal seed trajectories \added{in geometry N} for $\Ri_B=0$, black, $10^{-4}$, red, $10^{-3}$, blue, $3 \times 10^{-3}$, green and $10^{-2}$, purple.}
	\label{fig:energy1}
\end{figure}

We \replaced{investigate two}{use the} geometr\replaced{ies}{y}\added{, namely the `narrow' geometry `N'} investigated by \citet{Rabin2012} in the unstratified case, \deleted{namely} $4.35 \upi \times 2 \times 1.05 \upi$, one of the geometries considered by \citet{Butler1992} for linear, unstratified optimal perturbations\added{, in which the unstratified minimal seeds are localised in the streamwise direction but essentially fill the spanwise width, and $4.35 \upi \times 2 \times 2.10 \upi$ \added{ a `wide' geometry `W'}, which is twice as wide, and allows for localisation in the spanwise direction also. We used a modified version of the parallelised CFD solver Diablo \citep{Taylor2008} to solve the forward and adjoint equations, which uses Fourier modes in the streamwise $x$ and spanwise $z$ directions, and finite differences in the wall-normal $y$ direction, and uses a combined implicit-explicit Runge-Kutta-Wray Crank-Nicholson time integration scheme. The resolution for geometry N was $128 \times 256 \times 32$ and for geometry W was $128 \times 256 \times 64$.} 

Using the laddering down approach described above, we \replaced{have}{can} converge\added{d} to the minimal seed for $\Ri_B=0$ (confirming quantitatively the unstratified case of \citet{Rabin2012} using a different code, and objective functional $\mathcal{J}$), $\Ri_B = 10^{-4}$, $10^{-3}$, $3 \times 10^{-3}$ and $10^{-2}$ \added{in geometry N and $\Ri_B = 0$, $3\times 10^{-3}$ and $10^{-2}$ in geometry W, using the fixed value $T=300$ for all bulk Richardson numbers except for the largest, $\Ri_B = 10^{-2}$, for which we use $T=400$}.
The respective values of $E_c(\Ri_B)$ are shown in table \ref{tab:ec}. \added{We immediately see that $E_c$ is an increasing function of $\Ri_B$, as expected, since a stable stratification inhibits vertical motions, and so a transition process involving vertical motions should be expected to require a larger energy input. Interestingly, $E_c$ in geometry W is approximately 40\% of $E_c$ in geometry N for 
the same $Ri_B$. Since $E_c$, as defined in (\ref{eq:e0def}), is an energy density, and the volume
of geometry W is twice that of geometry N, this suggests that the minimal seeds in 
geometry W are spanwise localised,  and that the narrow geometry N actually
requires higher maximum amplitudes of perturbation in the minimal seed due to the enforced spanwise periodicity.}

\subsection{Minimal seeds in narrow geometry N}

\begin{figure}
	\centerline{
		\includegraphics[scale=0.81,trim=5cm 1cm 4cm 0cm]{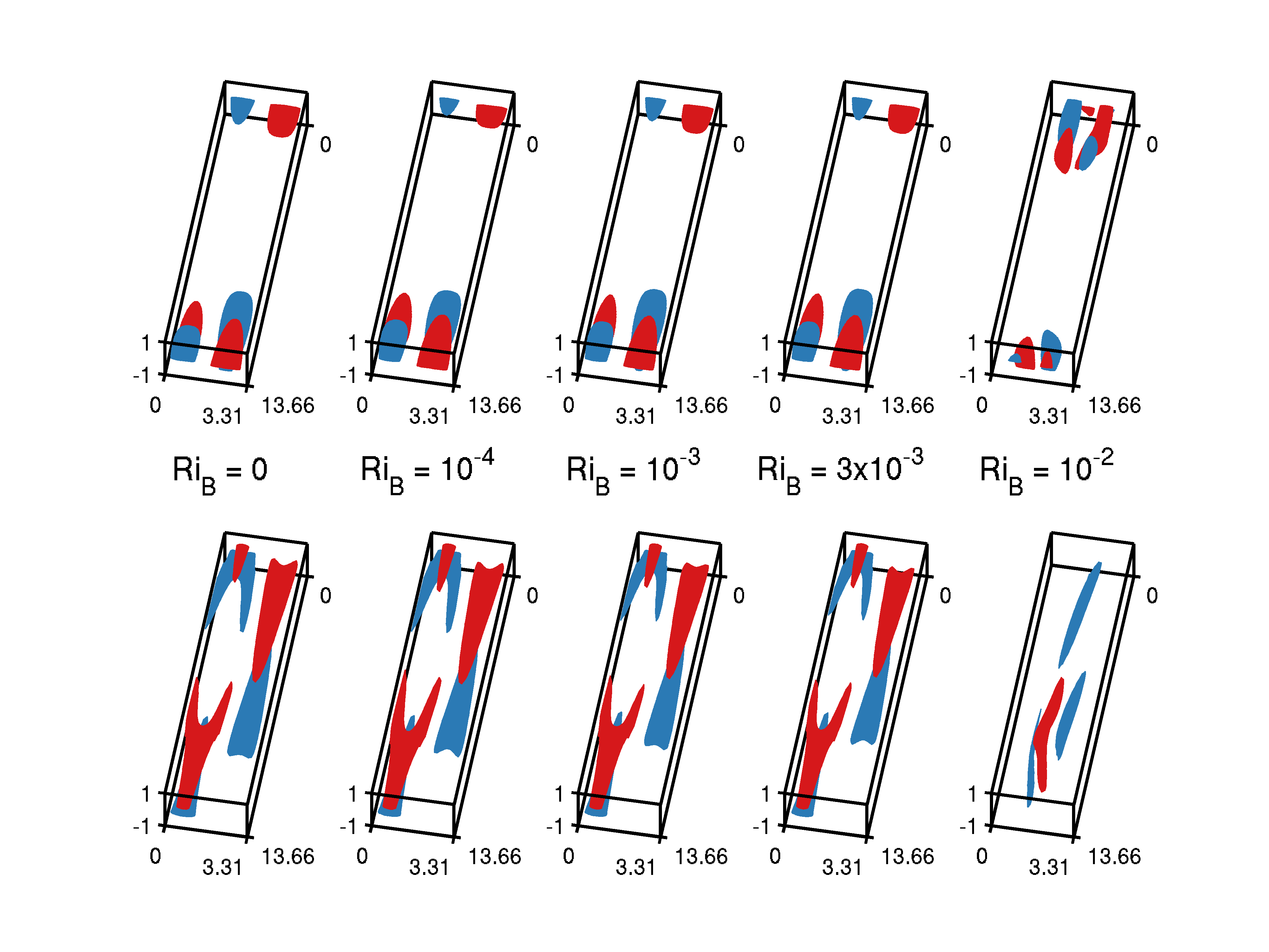}}
	\caption{Isosurfaces of streamwise perturbation velocity $u=\pm0.6 \max(u)$ at $t=0$ (top row) and $t=25$ (bottom row) for the minimal seed trajectories in geometry N for $\Ri_B=0$, $10^{-4}$, $10^{-3}$, $3 \times 10^{-3}$, and $10^{-2}$, from left to right. Videos
of the flow evolution are available as supplementary materials.}
	\label{fig:snapshota}
\end{figure}

\begin{figure}
	\centerline{
		\includegraphics[scale=0.81,trim=5cm 1cm 4cm 0cm]{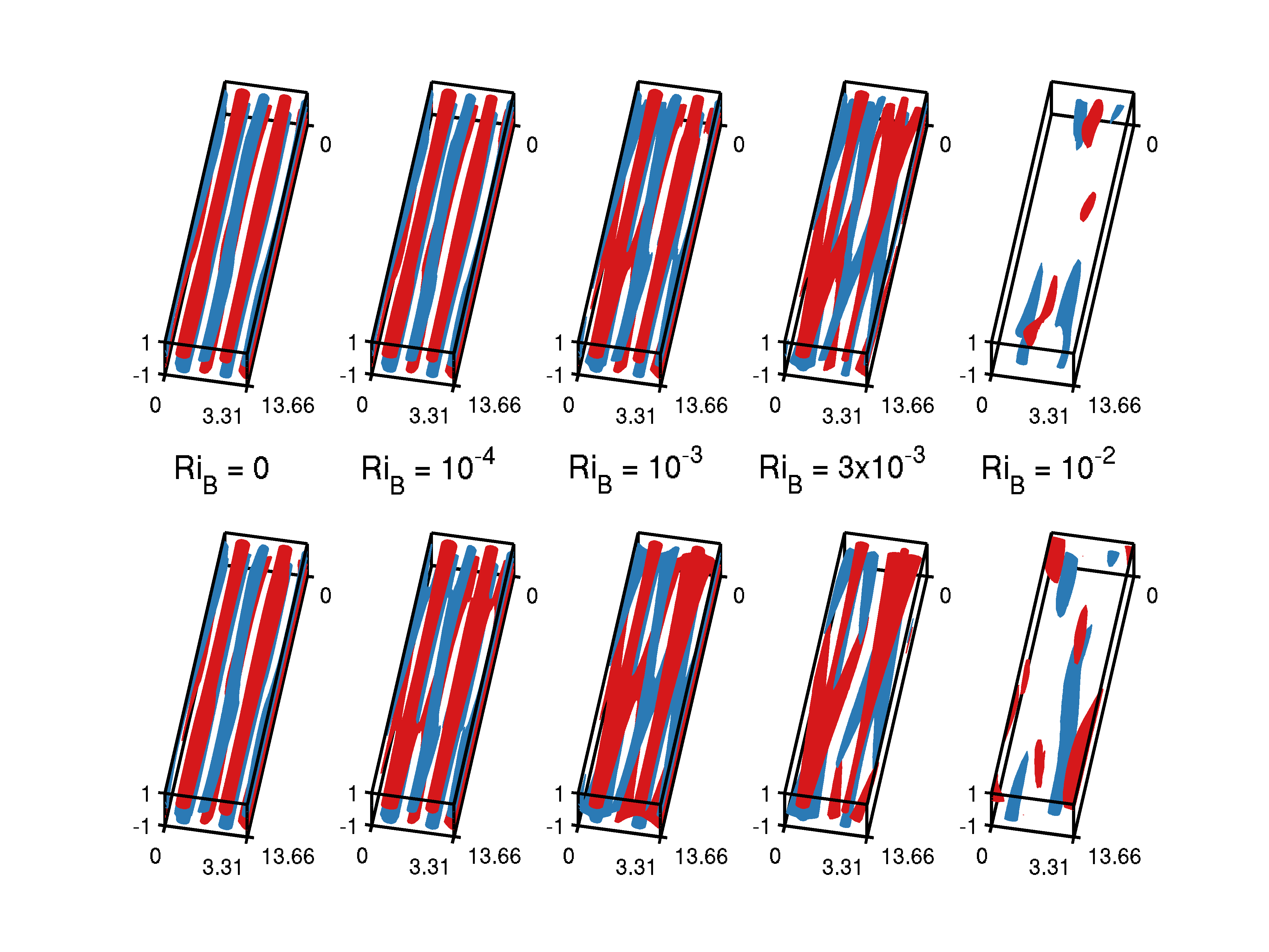}}
	\caption{Isosurfaces of streamwise perturbation velocity $u=\pm0.6 \max(u)$ at $t=70$ (top row) and $t=150$ (bottom row) for the minimal seed trajectories in geometry N for $\Ri_B=0$, $10^{-4}$, $10^{-3}$, $3 \times 10^{-3}$, and $10^{-2}$, from left to right. Videos
of the flow evolution are available as supplementary materials.}
	\label{fig:snapshotb}
\end{figure}

\begin{figure}
	\centerline{
		\includegraphics[scale=0.81,trim=5cm 1cm 4cm 0cm]{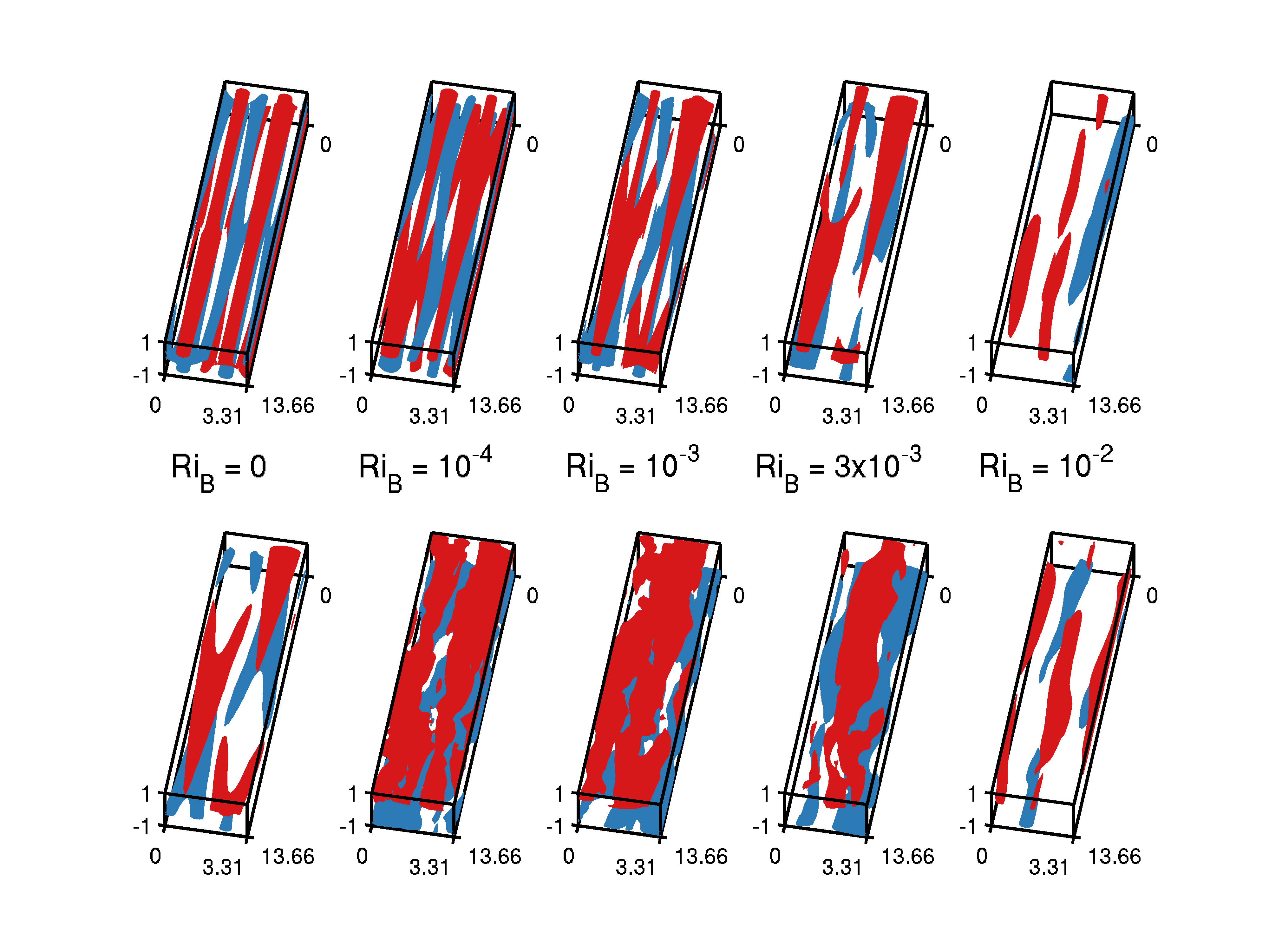}}
	\caption{Isosurfaces of streamwise perturbation velocity $u=\pm0.6 \max(u)$ at $t=210$ (top row) and $t=280$ (bottom row) for the minimal seed trajectories in geometry N for $\Ri_B=0$, $10^{-4}$, $10^{-3}$, $3 \times 10^{-3}$, and $10^{-2}$, from left to right. Videos
of the flow evolution are available as supplementary materials.}
	\label{fig:snapshotc}
\end{figure}

Figure \ref{fig:energy1} shows the time evolution from the \added{minimal seed} initial condition\added{s for geometry N} of the total energy density $E(t)$ defined as
\begin{equation}
E(t)= \frac{1}{2}\left< \boldsymbol{u} \bcdot \boldsymbol{u} + \Ri_B \rho^2 \right> = K(t)+P(t),\label{eq:edef}
\end{equation}
where $K(t)$ is the kinetic energy, $P(t)$ is the potential energy, and angled brackets 
denote volume averaging as defined in (\ref{eq:e0def}). \added{ Streamwise velocity $u=\pm0.6 \max(u)$ isosurfaces are plotted in figures \ref{fig:snapshota}, \ref{fig:snapshotb} and \ref{fig:snapshotc} at times $t=0$ and 25, $t=70$ and 150, and $t=210$ and 280 respectively,
and videos of the flow evolution are available as supplementary materials}.

The unstratified minimal seed, 
as shown in the leftmost column,  consists of a localised patch of flow structures aligned against the mean shear,
unwrapping via the Orr mechanism \citep[see][]{Orr1907} into an array of streamwise \replaced{aligned structures with a distinct oblique component, shown in the bottom left panel of  figure \ref{fig:snapshota}}{independent streaks}. \added{These oblique structures then transfer energy into streamwise independent streaks through the oblique wave mechanism in which structures with wavenumbers $(k_x,k_z) = (0,\pm a)$ interact nonlinearly to move energy into the wavenumeber $(k_x,k_z)=(0,0)$. }These streaks \replaced{are then able to `self-sustain' for a long period of the flow's evolution, as shown in the leftmost column of figure \ref{fig:snapshotb},}{then slowly increase in energy} by utilising the lift-up mechanism \citep[described by][]{Landahl1980} \added{and their own instability to offset viscous decay, \citep[see][]{Waleffe1997}}, before \added{eventually} being of large enough amplitude to \added{transition to a high energy oblique structure, as shown in the leftmost column of figure \ref{fig:snapshotc}, which is visited only transiently, leading to a} break down to small scale turbulence due to an
instability\deleted{associated with regions of high shear having been created, associated with a noticeable streamwise dependence, creating spanwise vortices} reminiscent of the Kelvin--Helmholtz instability. The energetics of these sequential growth mechanisms  in unstratified minimal seed trajectories were discussed by \citet{Pringle2012} and \citet{Duguet2013}.

This transition mechanism has been interpreted by \citet{Rabin2012} 
as an initial concentration of energy
in a small region of the flow, allowing a minimisation of the total energy input whilst maximising the local energy
input. From this point, the most efficient way to transition to turbulence is by exploiting the lift-up mechanism
on a high energy flow structure and thus increasing the total energy of the flow to a point beyond which it is sufficiently unstable to
become turbulent. As shown in figure \ref{fig:energy1}, there is a \replaced{sustained period of near-constant energy denstity $E(t)$, and during this time the flow is only very slowly changing, indicating that}{local minimum in the total energy density $E(t)$ at $t=110$, and at this time} the flow trajectory is \deleted{in a flat region of the edge manifold,} near a coherent state. \deleted{, remaining close to this state for a substantial period. In figure \ref{fig:snapshot} we plot isosurfaces of 60\% of the maximum positive and minimum negative streamwise velocity at $t=110$ for this minimal seed, showing that this coherent state is}\added{The isosurfaces plotted in the leftmost columns of figure \ref{fig:snapshotb} show that this state is} very reminiscent of an SSP/VWI state. As $\Ri_B$ is increased to $\Ri_B = 10^{-4}$ \added{and $10^{-3}$}, the minimal seed, its trajectory, and the coherent state to which it approaches, remain very similar to the unstratified case\added{, as can be seen from the flow isosurfaces
shown in the second and third columns of figures \ref{fig:snapshota}, \ref{fig:snapshotb} and \ref{fig:snapshotc}, although there are the beginnings of an oblique structure appearing in the coherent states to which these flows evolve}.

As $\Ri_B$ is increased further, the behaviour changes qualitatively. The minimal seed\deleted{s} for \deleted{$\Ri_B =10^{-3}$ and} $\Ri_B = 3 \times 10^{-3}$, (as shown in the fourth column, top row of figure \ref{fig:snapshota})  still consist\added{s} of a localised patch of flow structures aligned against the mean shear, which unwrap via the Orr mechanism \added{into the same array of streamwise aligned structures with a distinct oblique component, followed by the oblique wave mechanism. However, the newly created streamwise independent streaks are visited only transiently, no longer able to be sustained, and the flow evolves into a new, fully three dimensional coherent state which is shown in the fourth column of figure \ref{fig:snapshotb}}.\deleted{However, instead of unwrapping into an array of streamwise independent streaks, the flows unwrap directly onto the flow with high shear regions. These high shear regions are able to be sustained for a longer period of time. Whilst the mean shear is acting to intensify the high  shear regions, the high shear regions are tilted, and the density field is arranged such that it produces a torque in the opposite direction to the mean shear, allowing this structure to be maintained for a longer period of time. The critical energies $E_c$ of these minimal seeds are substantially larger than for the unstratified case. This is due to the apparent skipping of the lift-up phase during energy growth. Since stratification reduces the energy efficiency of vertical motions, the lift-up mechanism is no longer a viable growth mechanism, and so most of the energy growth must occur during the fast acting Orr mechanism, requiring the initial energies to become substantially larger. This switching off of the lift-up mechanism phase is consistent with a linear optimal perturbation analysis that we have performed, which showed that the linear optimal perturbation found in this geometry by 
for unstratified plane Couette flow consisting of streamwise independent streaks is no longer optimal for perturbation energy growth for $\Ri_B \gtrsim 7 \times 10^{-3}$, at which point three dimensional oblique structures utilising the Orr mechanism are optimal for linear transient perturbation energy growth at globally optimal times $T \lesssim 25$. In figure \ref{fig:snapshot} is also plotted isosurfaces of 60\% of the maximum positive and minimum negative perturbation streamwise velocity at the time of local minima in total perturbation energy density $t=88$ for the case $\Ri_B = 3 \times 10^{-3}$.} This new stratified coherent state is different from the unstratified SSP/VWI state \replaced{visited by the unstratified minimal seed trajectory}{shown in figure \ref{fig:snapshot}a}, with the stratified coherent state being clearly more three dimensional and very reminiscent of the stratified edge states recently reported by \citet{Olvera2014}.

Figure \ref{fig:energy1} shows that \added{in} the cases $\Ri_B = 10^{-4}$, $10^{-3}$ and $3 \times 10^{-3}$, transition to turbulence \added{is apparently} faster than the unstratified case, with the same accuracy in the estimated value of $E_c$\added{, and the same time target time $T=300$}. \added{However, the time taken to transition to turbulence is a function of how close our numerical estimates for the minimal seed initial conditions are to the edge manifold. In the limit of successively better approximations to the `ideal' minimal seed, which lies exactly on the edge manifold, $t_{transition} \rightarrow \infty$. Therefore, since the value of $E_c$ is only bracketed to a certain accuracy, the time to transition for the minimal seeds presented here is most likely to be affected by the actual value of $E_c$ relative to the energy bracket for $E_c$ found here.} \deleted{Stable stratification has not only affected the edge state and its stable manifold, but also its unstable manifold, and in particular the strength of the edge states' unstable directions appear to have increased significantly. Furthermore,} 

\replaced{T}{t}he minimal seed for $\Ri_B = 10^{-2}$ is once again of a qualitatively different character. Although it still consists of a localised patch of flow structures that unwrap via the Orr mechanism, there appears to be no quasi-steady flow structure \replaced{into}{onto} which it evolves.\deleted{There is a resemblance to the shear-baroclinic torque balance of the $\Ri_B=10^{-3}$ and $3 \times 10^{-3}$ minimal seeds, but} \replaced{T}{t}he flow is now chaotic with a weak oscillation, with no single structure dominating the flow evolution. 
 Indeed, due to the chaotic nature of the dynamics on the edge manifold that this trajectory follows, the DAL method struggled to identify turbulence solutions without extending the optimisation time interval to $T=400$. Even after extending the optimisation time interval, the DAL method required up to ten times as many iterations to identify initial conditions that transition to turbulence for the flow with $\Ri_B=10^{-2}$ compared to the number of iterations required
for flows with smaller bulk Richardson numbers,
apparently
because
of the chaotic nature of the edge manifold. As noted in the figure captions, videos of the evolution of the minimal seeds for the different values of $\Ri_B$ are available as supplementary material.

\subsection{`Wide' geometry W}

\begin{figure}
	\centerline{
		\includegraphics[scale=0.38,trim=2cm 0.5cm 1cm 0cm]{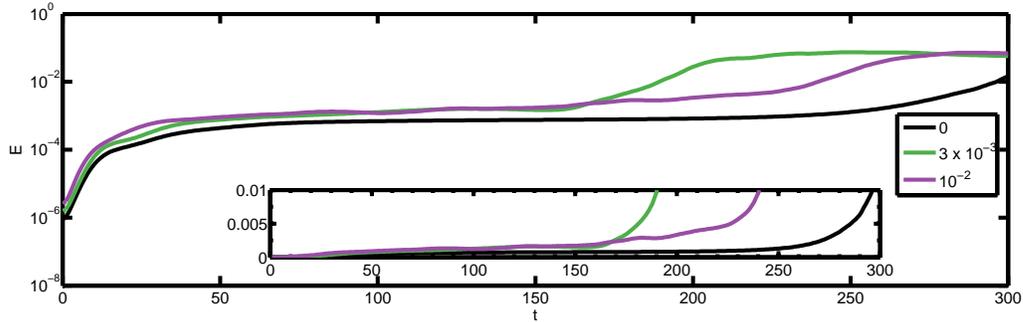}}
	\caption{Time variation of energy density $E(t)$ as defined in (\ref{eq:edef}) for the minimal seed trajectories \added{in geometry W} for $\Ri_B=0$, black, $3 \times 10^{-3}$, green, and $10^{-2}$, purple.}
	\label{fig:energy2}
\end{figure}

\begin{figure}
	\centerline{
		\includegraphics[scale=0.81,trim=5cm 1cm 4cm 0cm]{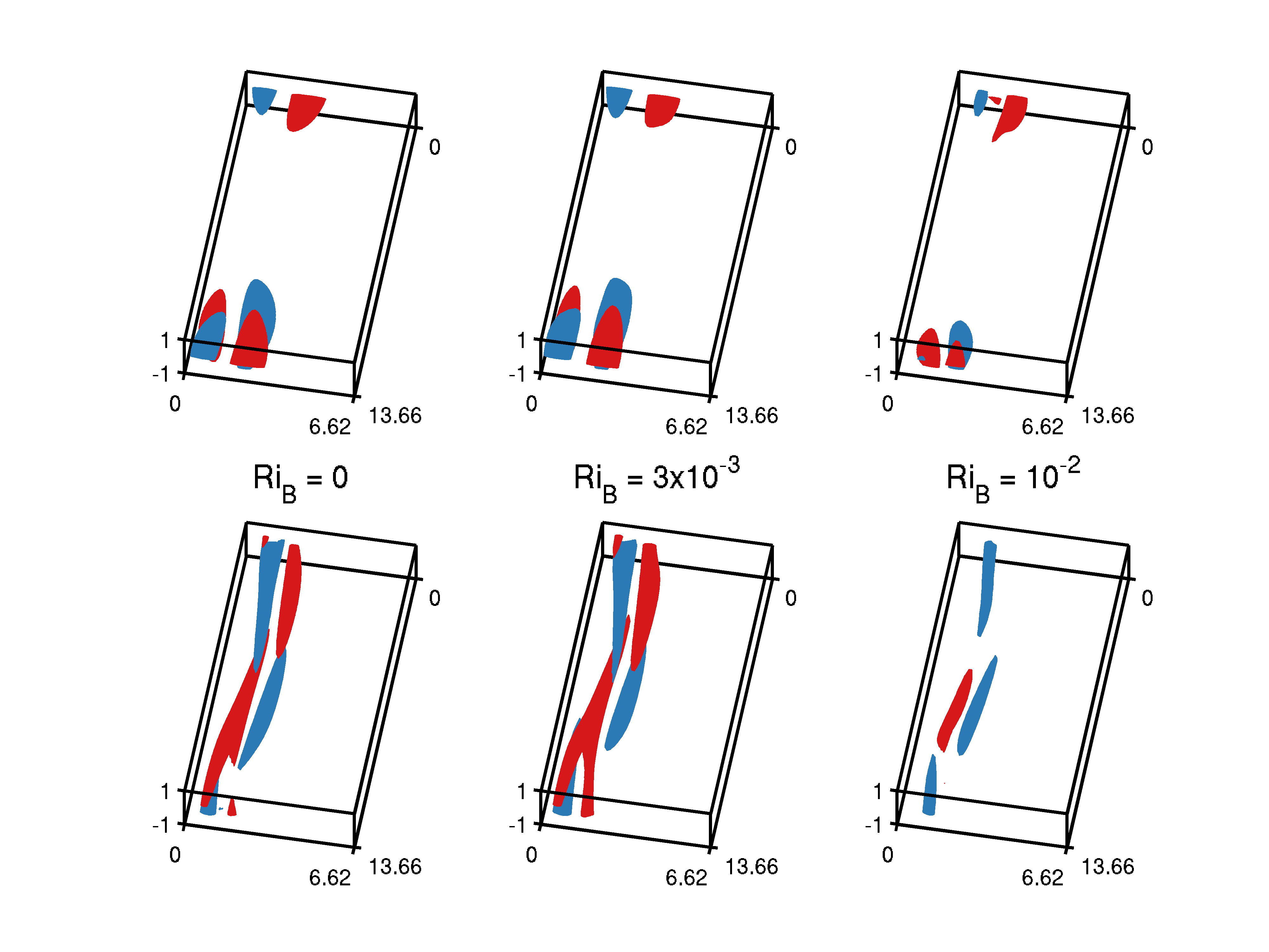}}
	\caption{Isosurfaces of streamwise perturbation velocity $u=\pm0.6 \max(u)$ at $t=0$ (top row) and $t=25$ (bottom row) for the minimal seed trajectories in geometry W for $\Ri_B=0$, $3 \times 10^{-3}$, and $10^{-2}$, from left to right.
Videos of the flow evolution are available as supplementary materials}
	\label{fig:snapshot2a}
\end{figure}

\begin{figure}
	\centerline{
		\includegraphics[scale=0.81,trim=5cm 1cm 4cm 0cm]{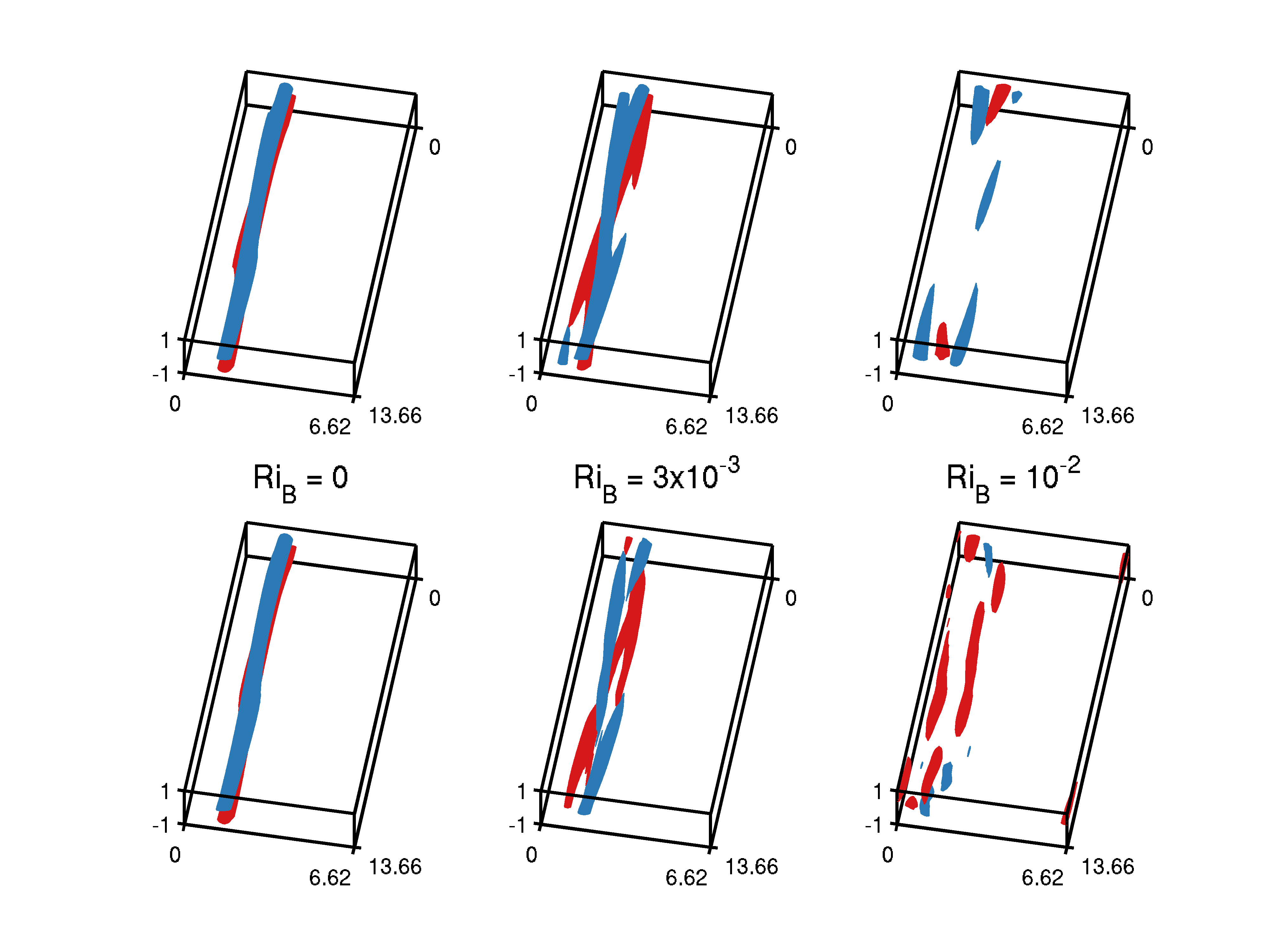}}
	\caption{Isosurfaces of streamwise perturbation velocity $u=\pm0.6 \max(u)$ at $t=70$ (top row) and $t=150$ (bottom row) for the minimal seed trajectories in geometry W for $\Ri_B=0$, $3 \times 10^{-3}$, and $10^{-2}$, from left to right. Videos
of the flow evolution are available as supplementary materials.}
	\label{fig:snapshot2b}
\end{figure}

\begin{figure}
	\centerline{
		\includegraphics[scale=0.81,trim=5cm 1cm 4cm 0cm]{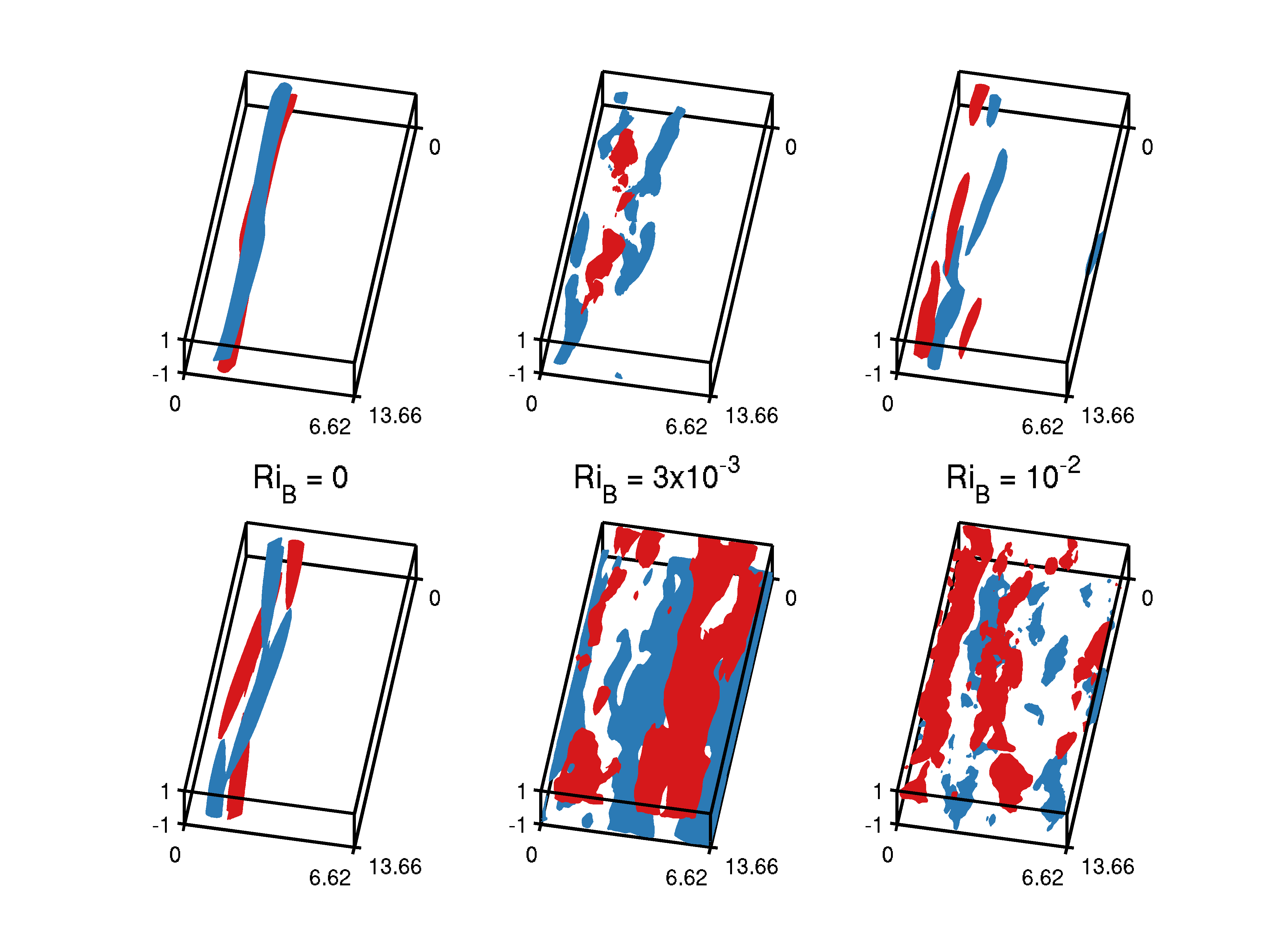}}
	\caption{Isosurfaces of streamwise perturbation velocity $u=\pm0.6 \max(u)$ at $t=210$ (top row) and $t=280$ (bottom row) for the minimal seed trajectories in geometry W for $\Ri_B=0$, $3 \times 10^{-3}$, and $10^{-2}$, from left to right. Videos
of the flow evolution are available as supplementary materials.}
	\label{fig:snapshot2c}
\end{figure}

As already noted, the minimal seeds in the narrow geometry N are streamwise localized, 
but fill much of the spanwise extent of the computational domain. To investigate to some extent
the sensitivity of the identified minimal seeds to the flow geometry, we also calculate minimal seeds in geometry W, which is twice as wide in the spanwise direction.
Figure \ref{fig:energy2} shows the time evolution from the minimal seed initial conditions of the total energy density
$
E(t)$ as defined in (\ref{eq:edef}) for the wide geometry W. 
 Streamwise velocity $u=\pm0.6 \max(u)$ isosurfaces are plotted in figures \ref{fig:snapshot2a}, \ref{fig:snapshot2b} and \ref{fig:snapshot2c} at times $t=0$ and 25, $t=70$ and 150, and $t=210$ and 280 respectively, and videos of the flow evolution are available as supplementary materials.

The $\Ri_B=0$ unstratified minimal seed trajectory in this wider geometry shares the same characteristic evolution as the minimal seed trajectory in the narrower geometry, but with the addition of spanwise localisation, that is a streamwise and spanwise localised patch of flow structures aligned against the mean shear which unwrap via the Orr mechanism into a streamwise aligned structure with a distinct oblique component, as shown in the leftmost column of figure \ref{fig:snapshot2a}, before utilising the oblique wave mechanism to produce a spanwise isolated pair of streaks,
reminiscent of the non-localised structure seen in geometry N. 
As is apparent in the tabulated values of $E_c$ listed in table \ref{tab:ec}, these structures
are slightly less energetic than the minimal seeds identified in the narrow geometry N, in that, as already noted, 
the critical values of the energy density $E_c$ (i.e. the energy divided by the volume of the computational domain) in geometry W are approximately 40\% of the equivalent values
determined in geometry N.

These streaks survive in the flow for an extended period of time, as shown in the leftmost column of  figure \ref{fig:snapshot2b}, and are another realisation of an SSP/VWI coherent structure. These streaks eventually transition to a high energy spanwise isolated oblique structure, as shown in the leftmost column of figure \ref{fig:snapshot2c}, which is visited only transiently, before breaking down to small scale turbulence. This sequence of events, as well as the spanwise localisation,  are both consistent with those reported by \citet{Monokrousos2011} and verified using a different objective functional by \citet{Rabin2012} in the domain $4\upi \times 2 \times 2\upi$ at the larger Reynolds number $\Rey = 1500$.

The minimal seed trajectory for $\Ri_B = 3 \times 10^{-3}$ (shown in the middle column of figures
\ref{fig:snapshot2a}, \ref{fig:snapshot2b} and \ref{fig:snapshot2c}) again shares the same characteristic evolution as the 
equivalent minimal seed trajectory in the narrow flow geometry N. The initial condition consists of the same spanwise and streamwise localised patch of flow structures aligned against the mean shear as in the unstratified flow, which again unwrap via the Orr mechanism into a streamwise aligned structure with a distinct oblique component, as shown in the lower row, middle column of figure \ref{fig:snapshot2a}. The oblique wave mechanism transfers energy into a pair of spanwise isolated streaks that are visited only transiently, and the flow evolves onto a long lived spanwise localised three dimensional coherent state which has the same oblique characteristics as the one found in the narrow geometry N. This structure is eventually no longer able to be maintained, and transition to turbulence occurs.

Furthermore, the minimal seed trajectory for $\Ri_B = 10^{-2}$ in geometry W is also a spanwise localised version of the equivalent trajectory in the 
narrower geometry N, as can be seen by comparison of the rightmost columns
of figures \ref{fig:snapshota}, \ref{fig:snapshotb} and \ref{fig:snapshotc} and
figures \ref{fig:snapshot2a}, \ref{fig:snapshot2b} and \ref{fig:snapshot2c}. The initial condition in geometry W consists of a streamwise and spanwise localised patch of flow structures that unwrap via the Orr mechanism into a flow which quickly becomes chaotic. The flow continues in much the same way as its small domain version, until eventually transitioning to turbulence. Once again, it appeared
computationally more difficult to identify the minimal seed for the 
flow with $Ri_B=10^{-2}$ than for the flows with 
smaller bulk Richardson numbers, 
requiring two or three times as many iterations, although this convergence
was still faster than for the equivalent flow
with the same $Ri_B=10^{-2}$ in the 
narrower geometry N.

\section{Rolls, streaks and waves}

To analyse the effect that stratification has on the SSP/VWI states, we decompose the full perturbation velocity and density fields into  roll, streak and wave \replaced{components. }{parts.} First, we define
$
\boldsymbol{\mathcal{U}} = \left< \boldsymbol{u} \right>_x$ and $
\Theta = \left< \rho \right>_x$,
where $ \left<\boldsymbol{a}\right>_x = \frac{1}{L_x} \int_0^{L_x} \boldsymbol{a} \,\, \mathrm{d}x$, and decompose
$
\boldsymbol{u} = \boldsymbol{\mathcal{U}} + \hat{\boldsymbol{u}}= 
\boldsymbol{\mathcal{U}}_r + \boldsymbol{\mathcal{U}}_s+\hat{\boldsymbol{u}}
$ and $\rho = \Theta + \hat{\rho}$,
so that
\begin{equation}
(\boldsymbol{u},\rho) = (\boldsymbol{\mathcal{U}}_r+\boldsymbol{\mathcal{U}}_s + \hat{\boldsymbol{u}},\rho)= (0,\mathcal{V},\mathcal{W},0)_{roll} + (\mathcal{U}, 0,0,\Theta)_{streak}+(\hat{u},\hat{v},\hat{w},\hat{\rho})_{wave}, \label{ssp}
\label{eq:urdef}
\end{equation}
where the $r$ subscript denotes the streamwise-independent wall-normal and spanwise
roll velocity and the $s$ subscript denotes the streamwise-independent streamwise streak velocity.
As originally argued independently by \cite{Hall1991} and \cite{Waleffe1997}, at high Reynolds number, if there are rolls in the flow of typical amplitude $\textit{O}(\epsilon)$ with $\epsilon \ll 1$, then their decay rate due to viscosity is $\textit{O}(\Rey^{-1})$. During the $\textit{O}(\Rey)$ time in which they survive in the flow, they can advect streamwise velocity through the $\textit{O}(1)$ shear of PCF a distance $\textit{O}(\epsilon \Rey)$ and so produce $\textit{O}(\epsilon \Rey)$ streaks in the flow. If the amplitude of these streaks is sufficiently large, they can undergo an instability which creates a wave field. The nonlinear self interaction of this wave field then puts energy back into the rolls, and provided that this input of energy is sufficient to balance the viscous decay of the rolls, a `self-sustaining process' associated with this `vortex-wave interaction' is possible. Supposing that the streaks need to be $\textit{O}(1)$ to become unstable requires $\epsilon = \Rey^{-1}$ and so a quadratic self-interaction of the wave field of $\textit{O}(\Rey^{-1})$ is sufficient. \citet{Hall2010} showed that there is \replaced{a numerically}{an} exact solution in unstratified PCF for asymptotically large $\Rey$ such that the rolls have $\mathcal{V},\mathcal{W}=\textit{O}(\Rey^{-1})$ and the streaks have $\mathcal{U}=\textit{O}(1)$ throughout the domain. The waves inject energy into the rolls primarily in an $\textit{O}(\Rey^{-1/3})$ critical layer where the background PCF plus the streak flow has the same velocity as the wave velocity, within which $(\hat{u},\hat{v},\hat{w}) = \textit{O}(\Rey^{-5/6},\Rey^{-7/6},\Rey^{-5/6})$, giving an integrated Reynolds stress contribution over the critical layer of $\textit{O}(\Rey^{-1})$. The waves essentially provide an $\textit{O}(\Rey^{-1})$ jump across the critical layer in the component of the roll velocity normal to the critical layer through their Reynolds stresses. Such scalings have been used successfully to provide initial guesses for a search for \added{numerically} exact solutions of the Navier--Stokes equations \citep[see][]{Waleffe2001,Wedin2004}.

To see how this scaling is affected by the addition of a stable stratification, we examine the energetics for the total energy density of the rolls and of the streaks. We obtain
\begin{eqnarray}
\frac{\mathrm{d}}{\mathrm{dt}} K_r(t)=\frac{1}{2}\frac{\mathrm{d}}{\mathrm{dt}} 
  \left< |\boldsymbol{\mathcal{U}}_r |^2 \right>_{y,z}  &=& \left< -\Ri_B \mathcal{V}\Theta
 -\boldsymbol{\mathcal{U}}_{r}\bcdot \left< \hat{\boldsymbol{u}}\bcdot \bnabla \hat{\boldsymbol{u}} \right>_x \right. \nonumber \\ & &
\left. \,\,\,\,\,\,\,\, - \Rey^{-1} \bnabla \boldsymbol{\mathcal{U}}_{r} \boldsymbol{:} \bnabla \boldsymbol{\mathcal{U}}_{r} \right>_{y,z}, \label{rolls} \\
\frac{\mathrm{d}}{\mathrm{dt}} K_s(t)=\frac{1}{2}\frac{\mathrm{d}}{\mathrm{dt}} 
  \left<   |\boldsymbol{\mathcal{U}}_{s} |^2 \right>_{y,z}  & = & \left< -\mathcal{V}\mathcal{U} 
 -\boldsymbol{\mathcal{U}}_{s}\bcdot \left< \hat{\boldsymbol{u}}\bcdot \bnabla \hat{\boldsymbol{u}} \right>_x  \right. \nonumber \\ & &
\left. \,\,\,\,\,\,\,\, -\Rey^{-1} \bnabla \boldsymbol{\mathcal{U}}_{s} \boldsymbol{:} \bnabla \boldsymbol{\mathcal{U}}_{s} \right>_{y,z}, \label{velstreaks} \\
\frac{\mathrm{d}}{\mathrm{dt}} P_s(t)=\frac{1}{2}\frac{\mathrm{d}}{\mathrm{dt}} 
\left<  \Ri_B \Theta^2 \right>_{y,z} & = &  \left<  \Ri_B \mathcal{V}\Theta 
-   \Ri_B \Theta \left< \hat{\boldsymbol{u}}\bcdot \bnabla \hat{\rho} \right>_x \right. \nonumber \\ & &
 \left . \,\,\,\,\,\,\,\,  -  \Ri_B (\Rey \Pran)^{-1} \bnabla \Theta \bcdot \bnabla \Theta \right>_{y,z}, \label{rhostreaks}
\end{eqnarray}
where $\left< \boldsymbol{a} \right>_{y,z} = \frac{1}{2L_z} \int_{-1}^1 \int_0^{L_z} \boldsymbol{a}\,\, \mathrm{d}z \mathrm{d}y$,  $K_r(t)$ is the roll kinetic 
energy density, $K_s(t)$ is the streak kinetic energy density, 
and $P_s(t)$ is the streak potential energy density.  Note that $K_r(t)+K_s(t) +K_w(t)= K(t)$,
where $K_w(t)$ is the wave kinetic energy density 
\begin{equation}
K_w(t)=\frac{1}{2} \langle | \hat{\mathbf{u}}|^2 \rangle. \label{kwdef}
\end{equation}

\begin{figure}
	\centerline{
		\includegraphics[scale=0.38,trim=2cm 0.5cm 1.8cm 0cm]{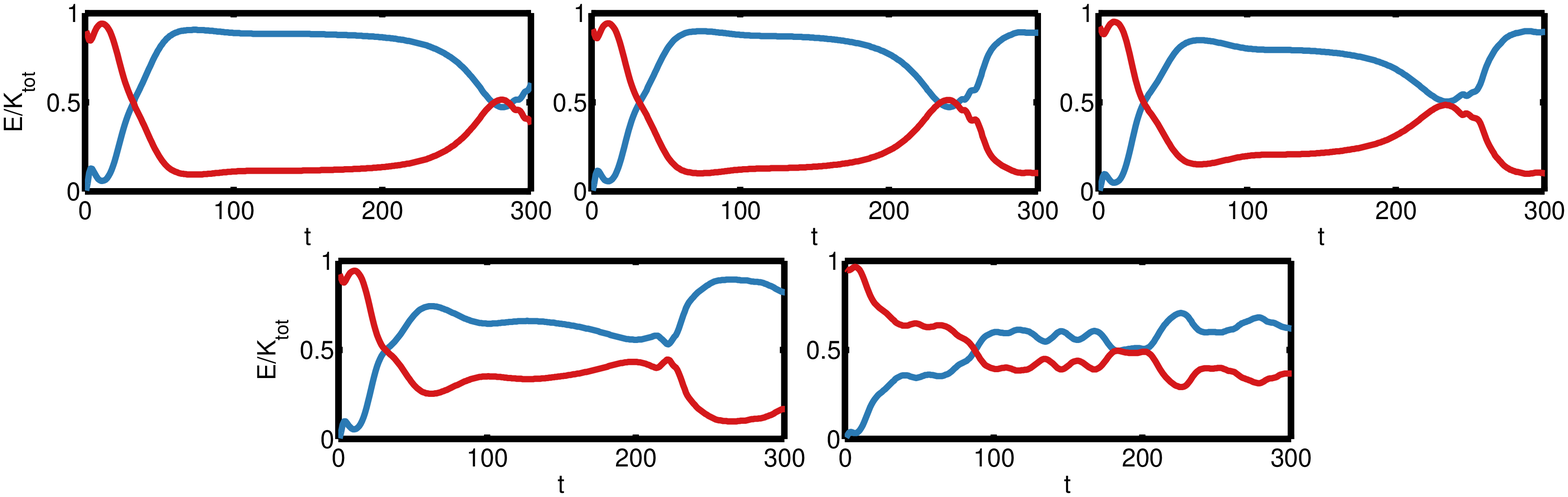}}
	\caption{Time dependence of  normalised streak kinetic energy density $K_s/K$ as
defined in (\ref{velstreaks}) (plotted with a blue line) 
and normalised wave kinetic energy density $K_w/K$ as defined in (\ref{kwdef}) (plotted with a red line)
for the minimal seed trajectories in geometry N for $\Ri_B = 0$ (upper left), $10^{-4}$ (upper middle), $10^{-3}$ (upper right), $3 \times 10^{-3}$ (lower left), $10^{-2}$ (lower right).}
	\label{fig:sspenergy1}
\end{figure}

\begin{figure}
	\centerline{
		\includegraphics[scale=0.38,trim=2cm 0.5cm 1.8cm 0cm]{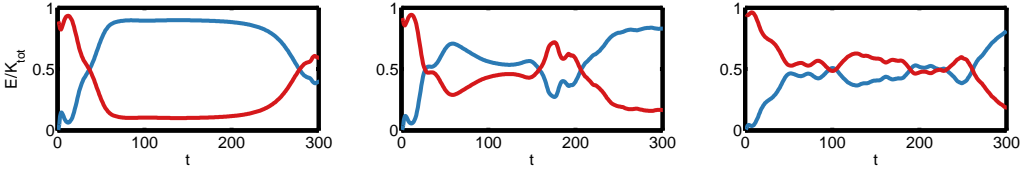}}
	\caption{Time dependence of  normalised streak kinetic energy density $K_s/K$ as
defined in (\ref{velstreaks}) (plotted with a blue line) 
and normalised wave kinetic energy density $K_w/K$ as defined in (\ref{kwdef}) (plotted with a red line)  for the minimal seed trajectories in geometry W for $\Ri_B = 0$ (left), $3\times 10^{-3}$ (middle) and $10^{-2}$ (right)}
	\label{fig:sspenergy2}
\end{figure}

Taking $\epsilon = \Rey^{-1}$,\deleted{we obtain} the correct scaling for the viscous decay time \added{is obtained} from (\ref{rolls}). We also see from (\ref{velstreaks}) that when $\mathcal{V}=\textit{O}(\Rey^{-1})$ is sustained over a period of $\textit{O}(\Rey)$, we obtain streaks $\mathcal{U}=\textit{O}(1)$. We  can also obtain the scaling of \citet{Hall2010} for the wave velocities, and their Reynolds stress contribution, under the assumption that they act over an $\textit{O}(\Rey^{-1/3})$ critical layer, and balance the viscous dissipation there.

For stratified PCF with $\Ri_B \neq 0$ there is a new buoyancy flux term $-\Ri_B \mathcal{V}\Theta$ entering the energetics of the rolls. Comparing the first term in the right-hand side of (\ref{velstreaks}) to that of (\ref{rhostreaks}), which are production terms for the streak  kinetic energy density and streak potential energy density respectively, we see that $\Theta = \textit{O}(1)$ everywhere for a passive scalar field placed in a self-sustained process. Thus, (\ref{rolls}) shows that a buoyancy flux associated with the streak flow effective across the whole domain provides a contribution to the kinetic energy density $K_r$ of the rolls of $\textit{O}(\Ri_B \Rey^{-1})$, and is able to disrupt fully the flux of wave Reynolds stress input of $\textit{O}(\Rey^{-3})$ over the $\textit{O}(\Rey^{-1/3})$ critical layer when $\Ri_B = \textit{O}(\Rey^{-2})$.\deleted{Recalling the the Reynolds number used for the minimal seed calculations above is $\Rey = 1000$,} \replaced{T}{t}his simple minded high Reynolds number scaling argument\added{, which is domain size independent,} is not inconsistent with the\deleted{$\Ri_B = 10^{-3}$ and $3 \times 10^{-3}$} moderate Reynolds number minimal seed calculations presented above for which the unstratified coherent state visited by the unstratified minimal seed trajectory is no longer a viable solution for sufficiently large $\Ri_B \sim 3 \times 10^{-3}$, and the new coherent states have a modified roll structure \added{that is no longer streamwise-independent.} \deleted{, relying on the density streaks to balance the shear production of the velocity streaks.}

Figures \ref{fig:sspenergy1} and  \ref{fig:sspenergy2} show the time evolution of 
the streak kinetic energy density $K_s(t)$, as 
defined in (\ref{velstreaks}) and plotted with a blue line,  
and the wave kinetic energy density $K_w(t)$, as 
defined in (\ref{kwdef}) and plotted with a red line,  
normalised by the total kinetic energy $K(t)$ for each of the minimal seed trajectories in geometries  N and W respectively. It is clear that for $\Ri_B=0$, $10^{-4}$ \added{and $10^{-3}$,} there is a balance (shown by the approximate plateaux in the streak and wave energy components) between these  energies (and also the residual roll kinetic density density $K_r(t)$, which is not shown as it is, unsurprisingly, appreciably smaller in magnitude) for a large period of the flow evolution. For $\Ri_B = 3\times 10^{-3}$, this balance has been significantly disrupted, and is not maintained purely by the velocity fields.\deleted{For $\Ri_B=3 \times 10^{-3}$,} \replaced{T}{t}here also is a small amplitude oscillation, and for $\Ri_B = 10^{-2}$ the above-mentioned weakly oscillatory nature of the flow is clear.\deleted{Also, \added{as shown in section 3,} the dynamics is clearly inherently three-dimensional, with} \replaced{T}{t}he unstratified self-sustaining process is completely disrupted for $\Ri_B \gtrsim 3 \times 10^{-3}$ as expected.

\section{Discussion}

Using the direct-adjoint looping (DAL) method, we have computed minimal seeds for turbulence (the initial conditions of smallest possible initial perturbation energy \added{density} $E_c(\Ri_B)$ that transition to turbulence) in stratified PCF for a range of bulk Richardson numbers $\Ri_B$ \added{in two geometries: a narrow geometry labelled N, which apparently allows only streamwise localisation of the minimal seed initial condition; and a twice as wide geometry W which also allows spanwise localisation}. In the unstratified case we converge to the same minimal seed found by \citet{Rabin2012} in geometry N, although here we use the time averaged dissipation rate objective functional in the DAL method instead of the total energy density at the target time\added{, and a minimal seed very similar to that of \citet{Monokrousos2011} in geometry W}. 

Since a stable stratification inhibits vertical motions, we see an increase of $E_c$ with $\Ri_B$, as expected. The minimal seeds follow trajectories in \replaced{state}{phase} space close to the edge manifold towards \replaced{a state in the edge manifold}{\replaced{an}{the} edge state}. These trajectories are found to spend a large amount of time in the vicinity of \replaced{such a}{\replaced{an }{the} edge} state\deleted{(or at least some other coherent structure \deleted{on or} nearby to the edge manifold)}. For unstratified flows, these coherent states are a realisation of an SSP/VWI state, and for sufficiently small $\Ri_B$ the coherent states are largely unchanged. For $\Ri_B = \textit{O}(\Rey^{-1})$ there is a slow draining of energy into the density field, creating a highly modified, yet still stationary coherent state, while for larger $\Ri_B$ the coherent states feature inherently three-dimensional chaotic motion with weak oscillations.

Examining the flow in terms of roll, streak and wave components as defined in (\ref{ssp}) demonstrates that the density field is expected, at asymptotically higher Reynolds number, to have a significant disrupting effect on the SSP/VWI process, and hence the coherent states, when $\Ri_B = \textit{O}(\Rey^{-2})$ through a removal of energy from vertical motion in the rolls, which is not entirely inconsistent with the minimal seed trajectories found at moderate Reynolds number. The effects of stratification on a wide class of exact coherent states in shear flows have yet to be studied in detail. We have demonstrated here that stratification disrupts the well-established high Reynolds number SSP/VWI states in PCF for the small value $\Ri_B = \textit{O}(\Rey^{-2})$ by affecting the energy input into the roll structures, the most delicate part of the interaction process, through an inhibition of vertical motions. For larger bulk Richardson numbers, we also observe chaotic solutions. We believe therefore that a careful re-examination of the SSP/VWI ansatz must be made for the case of stratified shear flows even with a very weak stratification when $\Rey \gg 1$, an important class of flows common in both environmental and industrial contexts.

\acknowledgements
Valuable discussions with Professor R. R. Kerswell, Professor Greg Chini, and Dr Sam Rabin are gratefully acknowledged. Thanks are also due to two anonymous referees, whose thoughtful and constructive comments have substantially improved this paper. T.S.E. is supported by a University of Cambridge SIMS Fund studentship. The research activity of C.P.C. is supported by EPSRC Programme Grant EP/K034529/1 entitled `Mathematical Underpinnings of Stratified Turbulence.' The data for the minimal seed initial conditions associated with this manuscript are made available at https://www.repository.cam.ac.uk /handle/1810/251121.

\bibliographystyle{jfm}
\bibliography{mybib}

\end{document}